\documentclass[%
  aps,%
  prl,%
  letterpaper,%
  superscriptaddress,%
  reprint,%
  longbibliography,%
]{revtex4-2}

\usepackage[preset=reset,pkgset=extended,hyperrefdefs={defer,noemail}]{phfnote}

\usepackage[thmset=rich,smallproofs=false,proofref={marginbottom}]{phfthm}

\usepackage[reset]{geometry}
\makeatletter\Gm@restore@org\makeatother

\usepackage{ragged2e}
\DeclareCaptionJustification{myjust}{\justify}
\usepackage[nohelv]{newtxtext}
\usepackage[varg]{newtxmath}

\usepackage{algorithm}
\usepackage{algpseudocode}

\usepackage[semibold]{sourcesanspro}

\usepackage{hyperref}
\usepackage[capitalize,nameinlink]{cleveref}

\makeatletter
\newcounter{maintheorem}
\phfMakeTheorem[counter=maintheorem,aliascounter=false,proofref=false]{maintheorem}{Theorem}

\def\paragraph#1{%
  \par\vspace{0.8ex plus 0.2ex minus 0.1ex}%
  \noindent\emph{#1\@addpunct{.}---}\,\ignorespaces}
\def\leq{\leqslant}
\def\geq{\geqslant}
\makeatother

\begin{document}
\title{Thermalization with partial information}
\author{Philippe Faist}
\affiliation{Dahlem Center for Complex Quantum Systems, Freie Universität Berlin, 14195 Berlin, Germany}
\author{Sumeet Khatri}
\affiliation{Department of Computer Science and Center for Quantum Information Science and Engineering, Virginia Tech, Blacksburg, VA 24061, USA}
\date{August 5, 2025}
\begin{abstract}
  A many-body system,
  whether in contact with a large environment or evolving under complex dynamics,
  can typically be modeled as
  occupying the thermal state singled out by Jaynes' maximum entropy principle.
  Here, we find analogous fundamental principles identifying a noisy quantum
  channel $\mathcal{T}$ to model the system's dynamics, going beyond the study
  of its final equilibrium state.
  Our maximum channel entropy principle states that $\mathcal{T}$ should
  maximize the channel's entropy, suitably defined, subject to any available
  macroscopic constraints.  These may correlate input and outputs, and may lead
  to restricted or partial thermalizing dynamics such as thermalization with
  average energy conservation.
  This principle is reinforced by an independent extension of the microcanonical
  derivation of the thermal state to channels, which leads to the same
  $\mathcal{T}$.
  Our technical contributions include a derivation of the general mathematical
  structure of $\mathcal{T}$, a custom postselection theorem relating an
  arbitrary permutation-invariant channel to nearby i.i.d.\@ channels, as well
  as novel typicality results for quantum channels for noncommuting constraints
  and arbitrary input states.
  We propose a learning algorithm for quantum channels based on the maximum
  channel entropy principle, demonstrating the broader relevance of
  $\mathcal{T}$ beyond thermodynamics and complex many-body systems.
\end{abstract}
\maketitle

The dynamics of quantum complex many-body systems have
seen a surge of recent interest~\cite{R0,R1,R2,R3,R4,R5,R6}, giving new momentum to the age-old question of how quantum systems
evolve towards thermal equilibrium~\cite{R7,R8,R9,R10,R11}.
Quantum chaotic dynamics has been associated with
scrambling and out-of-time-ordered
correlators~\cite{R12,R2,R13,R14,R15,R16,R17,R18,R19},
 operator entanglement~\cite{R20,R21,R22,R23,R24,R25,R26,R27},
energy level spacing statistics and random matrix theory~\cite{R28,R29,R30,R31,R13,R17,R32},
random unitary ensembles that form so-called
\emph{$k$-designs}~\cite{R13,R14,R33,R3,R34,R35,R36,R37,R38},
deep thermalization~\cite{R4,R5}, as well as
the long-time growth of quantum circuit
complexity~\cite{R0,R1,R3,R39,R38,R6}.
Past some initial relaxation time scale, such systems are typically modeled in their
canonical thermal state $\gamma$.  This model is justified through a variety of
standard arguments known from textbook statistical mechanics, including the
assumptions of ergodicity, of equipartition of microstates, or of weak contact
with a heat bath~\cite{R40,R41}, from Jaynes'
principle of maximum
entropy~\cite{R42,R43}, 
from canonical typicality~\cite{R7}, 
from the argument that accessible observables (or their time averages)
rapidly thermalize to their
thermal expectation values~\cite{R8,R9,R11},
as well as from the eigenstate thermalization
hypothesis~\cite{R44,R45,R46,R47,R48,R49,R37}.

\begin{figure}
  \centering
  \includegraphics{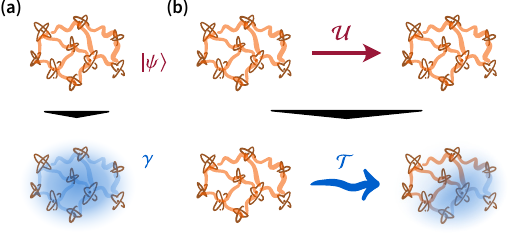}
  \caption{Thermal quantum state and thermal quantum channel as proxies for
    complex states and dynamics.  \textbf{\textsf{(a)}}~A quantum system in a
    state $\lvert {\psi}\rangle $ that is sufficiently complex (e.g., after a long-time
    chaotic evolution) is typically indistinguishable from the thermal state
    $\gamma$ for all practical purposes.  Jaynes' principle determines $\gamma$
    by maximizing the entropy over all states compatible with accessible
    expectation values.  \textbf{\textsf{(b)}}~In this work, we model complex
    dynamics $\mathcal{U}$ by some simpler noisy channel $\mathcal{T}$ that
    reproduces accessible expectation values of experiments that can be
    performed though choices of input states and output observables including a
    reference system.  Such constraints might mandate the channel $\mathcal{T}$
    to preserve information about its input, failing to fully thermalize the
    input system.  In this work, we study two natural prescriptions for
    determining $\mathcal{T}$ and prove that they coincide.}
  \label{z:bPuDarUIxB2T}
\end{figure}

Here, we ask whether a complex many-body quantum system's dynamics can be
modeled by a noisy quantum channel using similar fundamental principles as used
in the derivation of the thermal state.  For concreteness, suppose the dynamics
is some complex unitary evolution $\mathcal{U}$.  We seek a simpler, noisy
quantum channel $\mathcal{T}$ that reproduces accessible features of
$\mathcal{U}$, analogously to how the thermal state $\gamma$ can stand in as a
model for an unknown or complex pure state $\lvert {\psi}\rangle $
(\cref{z:bPuDarUIxB2T}).
The main result of this work is that two fundamental principles for deriving the
thermal state, the maximum entropy principle and the microcanonical
approach, have natural extensions to quantum channels, and both approaches lead
to the same \emph{thermal quantum channel} $\mathcal{T}$.

Our results are announced in this brief paper through high-level explanations;
our detailed mathematical derivations are the focus of our companion paper
ref.\@~\cite{R50}.

A starting point of our work is Jaynes' maximum entropy principle for quantum
states~\cite{R42,R43}.
Jaynes' principle asserts that the canonical thermal state maximizes the entropy
over all quantum states whose expectation values with respect to some set of accessible
macroscopic observables are fixed.  These observables are the macroscopically
controllable extensive degrees of freedom, such as energy and particle number.
The information-theoretic picture of Jaynes reveals a deeper role for the
canonical thermal state in information theory as the \emph{least informative},
or \emph{most uncertain} state compatible with prior information that is
encoded in a set of observables with fixed expectation
values~\cite{R42,R43,R51,R52,R53}.  Indeed, the thermal distribution is a central
concept in the mirror descent~\cite{R54} and matrix
multiplicative weights~\cite{R55,R56} algorithms.  The
canonical thermal state is centrally featured in algorithms for quantum
learning~\cite{R57,R58,R59,R60,R61,R62} and for semidefinite
programming~\cite{R63}.

We formulate a maximum entropy principle for quantum channels.  
We assume there is a set of accessible expectation values of experiments that
can be performed though choices of input states and output observables and which
may include a reference system.  Among the set of all quantum channels that
reproduce these expectation values exactly, we identify the channel that is the
most \emph{entropic}.  We employ a measure of channel entropy
defined and studied in
refs.\@~\cite{R64,R65,R66,R67},
and which has a natural interpretation suitable for modeling
thermalizing dynamics.
Specifically, a channel with high channel entropy always produces
highly entropic output states, regardless of its input state and even
conditioned on a reference system.
The channel $\mathcal{T}$ that maximizes the channel entropy subject to the
given constraints is called the \emph{thermal quantum channel}.

Our main technical contributions are to find the general mathematical structure
of thermal quantum channels, and to show that the thermal quantum channel
equivalently arises from global conservation laws in a larger, closed system.
This argument extends to channels the proof that the local reduced state of a
microcanonical ensemble is the canonical thermal state.
As a further result, we propose and numerically study
a learning algorithm for quantum channels based on our maximum channel entropy
principle; this algorithm extends similar techniques
for quantum states~\cite{R68,R69,R70,R71,R72,R73,R74,R59,R60}.

The robust theoretical foundations we lay for determining the thermal quantum
channel, using fundamental information-theoretic and physical principles,
supports its widespread usefulness from the description of partial, local, or
incomplete thermalizing dynamics to learning quantum channels.

This paper is structured as follows.  We first introduce our setting, recall the
definition of the channel entropy and formulate our maximum channel entropy
principle.  We then state our main results on the general mathematical structure
of thermal quantum channels and its equivalent derivation from the
microcanonical picture.  After some examples, we outline
our construction of a microcanonical channel, before finally
discussing our results.

\paragraph{Setting}
We consider a system $A$ along with a copy $R\simeq A$, which acts
as a reference system.
Some unknown, or complex, evolution $\mathcal{U}_{A\to B}$ maps states on $A$ to
some output $B$ (typically, $B$ and $A$ are the same system).
We further define the
canonical maximally entangled ket between $A$ and $R\simeq A$ as
$\lvert {\Phi_{A:R}}\rangle  = \sum_{j=1}^{d_A} \lvert {j}\rangle _A\otimes\lvert {j}\rangle _R$.
We assume that there are a set of physical properties of the system's dynamics
that are accessible to a macroscopic observer, and which should be reproduced
by $\mathcal{T}$.  For instance, we might be given pairs
$(\rho_A^{j}, Q_B^j)_{j=1}^J$ of input states and corresponding output
observables that we can prepare and measure, giving us access to the expectation
values $q_j \equiv \operatorname{tr}[{\mathcal{U}({\rho_A^j}) \, Q_B^j}]$.  
What is %
the ``least informative'' noisy quantum channel
$\mathcal{T}$ that is compatible with these expectation values?

These expectation values can capture correlations between the input and the
output of the channel, capturing the full quantum channel nature of the
system's evolution.
In fact, we consider more generally arbitrary linear constraints on
$\mathcal{T}$, written in terms of $\mathcal{T}$'s Choi matrix as
$q_j = \operatorname{tr}[{C^j_{BR} \, \mathcal{T}(\Phi_{A:R})}]$.
Input-output constraints mentioned above are expressed as
$C^j_{BR} = Q^j_B \otimes (\rho_A)^t$ with $(\cdot)^t$
denoting the transpose operation.

\begin{figure}
  \centering
  \includegraphics{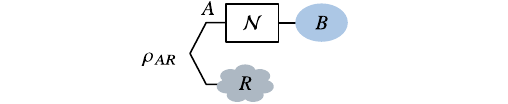}
  \caption{The entropy of a channel $\mathcal{N}_{A\to B}$ quantifies our
    minimal uncertainty about the output state of the channel, even if we keep a
    reference system $R$ as side information, and for all initial states
    $\rho_{AR}$ of the channel input and reference system.  Quantifying this
    uncertainty with the conditional quantum von Neumann entropy
    ${S}(B\mathclose{}\,|\,\mathopen{}R) = {S}(BR) - {S}(R)$, the entropy of a channel is given as
    ${S}(\mathcal{N}) = \min_{\rho_{AR}} {S}(B\mathclose{}\,|\,\mathopen{}R)_{\mathcal{N}(\rho_{AR})}$ %
    as per~\protect\cite{R64,R65,R66,R67}.  The entropy of a channel quantifies a property
    of always outputting highly entropic states, a common property of
    thermalizing dynamics.  The channel entropy is also the average
    thermodynamic work required to reset the output system to a fixed pure
    state, given access to $R$ and in the best case over input
    states~\protect\cite{R75,R76,R77}.}
  \label{z:85ylBc3ihN1H}
\end{figure}

We identify the ``least informative'' channel as the quantum channel
that minimizes the \emph{channel's entropy}~\cite{R64,R65,R66,R67} (\cref{z:85ylBc3ihN1H}).
The latter quantifies the uncertainty an observer always
has about the output of a
channel $\mathcal{N}_{A\to B}$, even if they access a reference system and 
can choose the input state of the channel.  The observer's uncertainty is
quantified using the conditional von Neumann entropy
${S}(B\mathclose{}\,|\,\mathopen{}R)_{\mathcal{N}(\rho_{AR})}$, defined as ${S}(B\mathclose{}\,|\,\mathopen{}R)_{\tau}
= -\operatorname{tr}[{\tau_{BR}\log(\tau_{BR})}] + \operatorname{tr}[{\tau_R\log(\tau_R)}]$.
This measure quantifies the average
resource requirements of many physical and information-theoretic tasks in the
presence of side information, such as thermodynamic information
erasure~\cite{R75,R76}, state
merging~\cite{R78}, and
decoupling~\cite{R79,R80}.
The entropy of the channel $\mathcal{N}_{A\to B}$ is defined
as~\cite{R64,R65,R66,R67}:
\begin{align}
  {S}(\mathcal{N}_{A\to B})
  = \min_{\rho_{AR}} {S}(B\mathclose{}\,|\,\mathopen{}R)_{\mathcal{N}(\rho_{AR})}\ .
\end{align}
Because ${S}(B\mathclose{}\,|\,\mathopen{}R)_{\tau}$ is concave in $\tau_{BR}$, the minimum is
necessarily achieved by some pure state $\lvert {\phi}\rangle _{AR}$.  All pure states
$\lvert {\phi}\rangle _{AR}$ on $AR$, up to a unitary on $R$, can be parametrized as
$\phi_R^{1/2}\,\lvert {\Phi_{A:R}}\rangle $ where $\phi_R$ is a density matrix.
Henceforth, we replace the minimization over
$\rho_{AR}$ by a minimization over $\phi_R$ in this fashion.

\noindent\textit{\textbf{Maximum channel entropy principle:}
As a ``least informative'' estimate for the unknown or complex $\mathcal{U}$, we
choose the quantum channel $\mathcal{T}$ that maximizes ${S}(\mathcal{T})$
subject to the constraints $\operatorname{tr}[{ C^j_{BR} \, \mathcal{T}(\Phi_{A:R}) }] = q_j$
for $j=1, \ldots, J$.}

We call a channel $\mathcal{T}$ that satisfies the maximum channel entropy
principle a \emph{thermal quantum channel}.  If the maximizer is not unique,
we choose for the purposes of this short paper a $\mathcal{T}$ which is a
limit of channels $\{{\mathcal{T}^\epsilon}\}$ that maximize
${S}(B\mathclose{}\,|\,\mathopen{}R)_{{\mathcal{T}^\epsilon(\phi_{AR}^\epsilon)}}$ for
full-rank $\phi_R^\epsilon$ with $\phi_R^\epsilon\to\phi_R$.

\paragraph{Main technical result}
Our core technical contribution is to prove that the quantum thermal channel,
defined via the maximum channel entropy principle, can be equivalently derived
through independent physical considerations based on conservation laws on a
larger system (microcanonical picture).  Our main result is broken up into the
following parts.

First, we find the general mathematical structure of any channel that is
obtained from the channel maximum entropy principle.
\begin{maintheorem}
  \label{z:na-CDYw1CUzm}
  Fix a set of constraints $\operatorname{tr}[{C^j_{BR} \mathcal{T}(\Phi_{A:R})}] = q_j$ for
  $j=1, \ldots, J$.  A quantum channel $\mathcal{T}$ is a quantum thermal
  channel if and only if it satisfies all constraints and is of the form
  \begin{align}
    \mathcal{T}(\Phi_{A:R})
    = \phi_R^{-1/2} {e}^{ \phi_R^{-1/2}
    \bigl({\mathds{1}_B\otimes \bar{F}_R - \sum_{j=1}^J \mu_j C^j_{BR} }\bigr)
    \phi_R^{-1/2} } \phi_R^{-1/2}\ ,
    \label{z:cTWuXXQ4W6pj}
  \end{align}
  where $\mu_j\in\mathbb{R}$, $\bar{F}_R$ is a Hermitian matrix, and where
  $\lvert {\phi}\rangle _{AR}=\phi_R^{1/2}\lvert {\Phi_{A:R}}\rangle $ is optimal in
  ${S}(\mathcal{T})$;
  if $\phi_R$ is rank-deficient, then~\eqref{z:cTWuXXQ4W6pj} is to
  be understood as a limit of channels of this form
  for full-rank $\phi_R^\epsilon$ with $\phi_R^\epsilon\to\phi_R$.
\end{maintheorem}

The general form~\eqref{z:cTWuXXQ4W6pj} extends the familiar Gibbs
canonical form of the thermal state ${e}^{-\beta H}/Z$, where $H$ is the system
Hamiltonian, $\beta$ the inverse temperature, and $Z$ the partition function.
In \cref{z:na-CDYw1CUzm},
the real numbers $\mu_j$ are ``generalized chemical potentials'' introduced as
Lagrange dual variables associated with each constraint in the maximum channel
entropy problem.  The $\bar F_R$ matrix is the Lagrange dual variable associated
with the trace-preserving constraint on $\mathcal{T}$; it generalizes the free
energy of a thermal state.
The proof of \cref{z:na-CDYw1CUzm} relies on tools from convex
optimization and Lagrange duality~\cite{R81} (cf.\@
ref.~\cite{R50}).  It exploits the fact that the minimum
over $\phi_R$ and the maximum over the channel can be
interchanged~\cite{R82,R83}.

The second part of our main result is to identify a
microcanonical channel by imposing conservation laws implied by the constraints
on many copies of the system.  We consider a large number $n$ of copies of the
$AR$ systems.  The first copy acts as the system of interest, and the remaining
$n-1$ copies can be thought of as a large environment or bath.  We consider a
global process $\mathcal{E}_{A^n\to B^n}$ that describes the evolution of the
system along with its large environment.
We formalize conservation laws for processes by requiring that any experiment
that estimates the constraint expectation value using any arbitrary full-rank
input state produces sharp statistics around the constraint value $q_j$ in the
limit of large $n$.  We explain this formalization in more detail
below.
Among all global channels that obey these conservation laws,
we identify the channel with
the maximum entropy as the \emph{microcanonical channel} $\Omega_{A^n\to B^n}$.
The latter leads to the thermal quantum channel:
\begin{maintheorem}
  \label{z:0F-2ltPZrRc9}
  Let $\Omega_{A^n\to B^n}$ be a microcanonical channel for the conservation
  laws induced by the constraint values $q_j$.  Let $\mathcal{T}_{A\to B}$ be
  the single-copy quantum thermal channel with constraint values $q_j$.
  Let $\lvert {\phi}\rangle _{AR} = \phi_R^{1/2}\lvert {\Phi_{A:R}}\rangle $ be the state that is
  optimal in the channel entropy ${S}(\mathcal{T})$.  Then
  \begin{align}
    \operatorname{tr}_{n-1}[{ \Omega_{A^n\to B^n}({\phi_{AR}^{\otimes n}})}]
    \approx
    \mathcal{T}(\phi_{AR})\ .
    \label{z:hc5i2B-Sc8xb}
  \end{align}
  If $\phi_R$ is rank-deficient, then this statement is to be understood in the
  limit of large $n$ and for full-rank states $\phi^{\epsilon}_R\to\phi_R$.
\end{maintheorem}

\Cref{z:0F-2ltPZrRc9} completes our main result by proving
that the microcanonical channel on $n$ copies of the system, when looking at its
effective action on a single system, reproduces the thermal quantum channel
obtained via the maximum channel entropy principle.  Since $\lvert {\phi}\rangle _{AR}$ is a
pure state with full-rank reduced states, the proximity of the single-copy
resulting state on $BR$ to $\mathcal{T}(\phi_{AR})$ ensures that the process
induced by $\Omega$ onto the first copy $A\to B$, using $\phi_{AR}$ as inputs to
all the $n-1$ environment systems, is itself close to $\mathcal{T}$ as a quantum
process.  This proximity could be quantified in terms of the diamond norm,
while picking up an additional constant factor of $d_A$.

\paragraph{Examples of thermal quantum channels}
We compute the quantum thermal channel associated with the following example
sets of constraints (cf.\@ Appendix and ref.\@~\cite{R50}
for details).

If no constraints are imposed at all, the thermal quantum channel is the
completely depolarizing channel
$\mathcal{D}_{A\to B}(\cdot) = \operatorname{tr}(\cdot)\, \mathds{1}_B/d_B$, where $d_B$ is the
Hilbert space dimension of $B$.

Now consider a single constraint on the output system, taken to be of the form
$C^{1}_{BR} \equiv H_B \otimes (\rho_A)^t$ for some output observable $H_B$ and
arbitrary input state $\rho_A$, and fix $q_1\in\mathbb{R}$.  We find that the
associated thermal quantum channel replaces its input by the output Gibbs state:
$\mathcal{T}(\cdot) = \operatorname{tr}({\cdot})\,{e}^{-\beta H_B}/Z$, where $\beta, Z$ are
determined by the constraint value $q_1$.  Interestingly, we find the same
channel regardless of the input state $\rho_A$ used to impose the constraint.
We find the same channel even if we impose this constraint for all input states.
(This can be done with a finite set of constraints by finding a tomographically
complete set of input states.)

If we impose our channel to strictly conserve energy, i.e., to map states
supported on an energy eigenspace to states supported on the same energy
eigenspace, we find a channel $\mathcal{T}(\cdot)$ that applies the completely
depolarizing channel within each energy eigenspace.  

We can now demand of our channel that it conserves the average energy of a
state, for given Hamiltonians $H_A$ and $H_B$.  I.e., we demand that
$\operatorname{tr}[{\mathcal{T}(\rho_A) H_B}] - \operatorname{tr}({\rho_A H_A}) = 0$ for all states $\rho_A$.
This condition can be imposed with a finite set of linear constraint operators
$C^j_{BR} = H_B\otimes (\rho^j_A)^t - \mathds{1}_B\otimes [{(\rho_A^j)^{1/2} H_A
(\rho_A^{j})^{1/2}}]^t$ with $q_j=0$ for a finite set of states $\{{ \rho^j_A }\}$
whose linear span includes all density matrices.  In this situation we find
\begin{align}
  \mathcal{T}(\cdot) = \sum_E \langle {E}\mkern 1.5mu\relax \vert \mkern 1.5mu\relax {\cdot}\mkern 1.5mu\relax \vert \mkern 1.5mu\relax {E}\rangle _A 
  \, \frac{{e}^{-\beta(E)\, H_B}}{Z(E)}\ ,
  \label{z:IILP0BFRJrWc}
\end{align}
where $\lvert {E}\rangle _A$ are the eigenstates of $H_A$, and where $\beta(E), Z(E)$ are
the inverse temperature and partition function of a canonical Gibbs state for $H_B$ 
at energy $E$.  %
In other words, the thermal channel~\eqref{z:IILP0BFRJrWc}
measures the input energy and prepares the output thermal state that is
compatible with the measured value of energy.
This channel describes thermalizing dynamics, in that it always outputs a
thermal state.  Yet, it conserves memory of its input state, since the output
state's average energy coincides with the input state's.
This $\mathcal{T}$ is an example of a channel that is necessarily obtained by
applying the maximum channel entropy principle on the dynamics themselves,
rather than considering properties of the output state alone.  More
specifically, should the input state to $\mathcal{T}$ be a mixture between
several energy levels, the output is a corresponding mixture of Gibbs states.
This output state differs from the canonical state at the input state's average
energy, which a naive application of Jaynes' maximum entropy principle would
have led us to conclude.

Further example situations include classical channels, where we recover some
existing results~\cite{R84,R85},
and Pauli channels, for which the thermal channel's Choi
state is thermal in the Bell basis.

\begin{figure}
   \centering
   \includegraphics{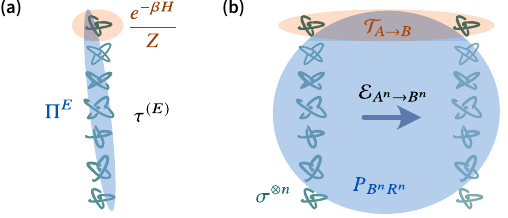}
   \caption{Construction of the microcanonical channel.
     \textsf{\textbf{(a)}}~The canonical thermal state ${e}^{-\beta H}/Z$ arises
     as the reduced state of a microcanonical state $\tau^{(E)}$ supported on
     the microcanonical subspace with projector $\Pi^E$, typically justified via
     global energy conservation.  \textsf{\textbf{(b)}}~We construct an operator
     $P_{B^nR^n}$ (analogous to $\Pi^E$) that selects all quantum channels
     $\mathcal{E}_{A^n\to B^n}$ that have sharp statistics when the expectation
     value constraints are estimated using any input state $\sigma^{\otimes n}$.
     ($R^n$ are reference systems which are not depicted.)  By `$P_{B^nR^n}$
     selects $\mathcal{E}$' we mean
     $\operatorname{tr}[{P_{B^nR^n} \mathcal{E}(\sigma^{\otimes n})}] \approx 1$ for all
     $\sigma$.  The microcanonical channel is identified, analogously to the
     microcanonical state, as the most entropic channel that is selected by
     $P_{B^nR^n}$.}
   \label{z:q2wrp8XV54Q1}
 \end{figure}

\paragraph{Construction of the microcanonical channel}
We extend the derivation of the maximum-entropy canonical state from the
microcanonical ensemble of refs.~\cite{R86,R87}.  We replace the single-copy expectation value
constraint $\operatorname{tr}[{C^j_{BR} \mathcal{T}(\Phi_{A:R})}] = q_j$ by a global
conservation law on the $n$ systems, representing a quantum experiment that
measures the global constraint value (\cref{z:q2wrp8XV54Q1}).
The most general single-copy experiment
to test the constraint
involves preparing a pure state $\lvert {\sigma}\rangle _{AR}$, applying the channel
$\mathcal{T}_{A\to B}$, and measuring some observable $H_{BR}$ on the joint
system $BR$.  Any such input pure state can be written as
$\lvert {\sigma}\rangle _{AR} = \sigma_R^{1/2}\lvert {\Phi_{A:R}}\rangle $ up to a unitary on $R$; the
latter can be absorbed into the observable $H_{BR}$.  Fixing any full-rank
$\sigma_R$ and defining
$H^{j,\sigma}_{BR} \equiv \sigma_R^{-1/2} C_{BR}^j \sigma_R^{-1/2}$, we find the
expected measurement outcome is the desired constraint value:
$\operatorname{tr}[{H^{j,\sigma}_{BR} \mathcal{T}(\sigma_{AR})}] = \operatorname{tr}[{H^{j,\sigma}_{BR}
\sigma_R^{1/2} \mathcal{T}(\Phi_{A:R}) \sigma_R^{1/2}}] = \operatorname{tr}[{C^j_{BR}
\mathcal{T}(\Phi_{A:R})}]$.
Now, we perform this measurement on $n$ copies of the systems $AR$ after the
joint application of a global process $\mathcal{E}_{A^n\to B^n}$, and we compute
a statistical average of the outcomes.  We suppose that the input state is the
independent and identically distributed (i.i.d.\@) state
$\sigma_{AR}^{\otimes n}$.  This procedure is equivalent to measuring the
observable
$\overline{H^{j,\sigma}}_{B^nR^n} = \frac1n \sum_{i=1}^n \mathds{1}_{BR}^{\otimes
  i-1} \otimes H^{j,\sigma}_{B_iR_i} \otimes \mathds{1}_{BR}^{\otimes (n-i-1)}$ on
the state $\mathcal{E}_{A^n\to B^n}(\sigma_{AR}^{\otimes n})$.
For any $\eta>0$, denote by
$\{{ \overline{H^{j,\sigma}}_{BR} \in [q_j \pm \eta] }\}$ the projector onto the
eigenspaces of $\overline{H^{j,\sigma}}_{B^nR^n}$ associated with eigenvalues in
the interval $[q_j-\eta,q_j+\eta]$.
We say that a quantum channel $\mathcal{E}_{A^n\to B^n}$ \emph{has sharp
  statistics for $\overline{H^{j,\sigma}}_{B^nR^n}$ around $q_j$} if
$\operatorname{tr}[{ \{{ \overline{H^{j,\sigma}}_{BR} \in [q_j \pm \eta] }\} \,
\mathcal{E}_{A^n\to B^n}(\sigma_{AR})}] \approx 1$.

The operator $C^j_{BR}$ holds no inherent information about what input state
should be used to test the constraint.  To capture the channel nature of the
problem, we would like a microcanonical channel to have sharp statistics for
$\overline{H^{j,\sigma}}_{B^nR^n}$ around $q_j$ for all input states $\sigma_R$.
However, the operator $H^{j,\sigma}_{BR}$ might have a diverging norm if
$\sigma_R$ has minuscule eigenvalues.  This property could prevent convergence
of the outcome statistics of $n$-copy sample average measurement of
$H^{j,\sigma}_{BR}$: The latter might have huge eigenvalues that appear with
vanishing probability but contribute meaningfully to the constraint expectation
value.  %
To remedy this problem, we demand from a microcanonical channel to have sharp
statistics for $\overline{H^{j,\sigma}}_{B^nR^n}$ around $q_j$ for all input
states $\sigma_R$ with all eigenvalues above some fixed threshold $y>0$ (i.e.,
$\sigma_R \geq y\mathds{1}$).  As we increase $n$, we can correspondingly decrease
$y$ so as to ensure that for $n\to\infty$, the constraint is tested for all
input states.

The following theorem formalizes our generalization of the microcanonical
subspace to quantum channels (see Appendix and
ref.~\cite{R50} for specific error tolerance parameters).
\begin{maintheorem}
  \label{z:EdbnDu7-QZV8}
  There exists $0\leq P_{B^nR^n}\leq \mathds{1}$ such that both following conditions
  hold:
  \begin{enumerate}[(\roman*)]
  \item Let $\mathcal{E}_{A^n\to B^n}$ be any quantum channel that obeys
    $\operatorname{tr}[{ P_{B^nR^n} \mathcal{E}(\sigma_{AR})}]\approx 1$ for all $\sigma_R>y\mathds{1}$.
    Then $\mathcal{E}_{A^n\to B^n}$ has sharp statistics for
    $\overline{H^{j,\sigma}}_{B^nR^n}$ around $q_j$ for all $\sigma_R>2y\mathds{1}$.
  \item Let $\mathcal{E}_{A^n\to B^n}$ be any quantum channel that has sharp
    statistics for $\overline{H^{j,\sigma}}_{B^nR^n}$ around $q_j$ for all
    $\sigma_R>y\mathds{1}$.  Then $\operatorname{tr}[{ P_{B^nR^n} \mathcal{E}(\sigma_{AR})}]\approx 1$
    for all $\sigma_R>2y\mathds{1}$.
  \end{enumerate}
\end{maintheorem}

Finally, we leverage our $P_{B^nR^n}$ to identify a microcanonical channel.  The
microcanonical quantum state is identified as the state with support in the
microcanonical subspace which has uniform spectrum, equivalently which is a Haar
average of all states in the microcanonical subspace, and which is the most
entropic.  Here, we identify the \emph{microcanonical channel
  $\Omega_{A^n\to B^n}$ associated with $P_{B^nR^n}$} as the most entropic
channel which satisfies $\operatorname{tr}[{\Omega_{A^n\to B^n} P_{B^nR^n} }] \approx 1$
(specific error parameters are detailed in the Appendix and
ref.~\cite{R50}).  From this channel, we
can recover the thermal quantum channel $\mathcal{T}$, defined via the maximum
entropy channel principle, through \cref{z:0F-2ltPZrRc9}.

\paragraph{Inference theory and an algorithm for learning quantum channels} A prominent application of the maximum-entropy principle for quantum states is in the reconstruction of quantum states using %
incomplete %
knowledge, in the form of expectation-value estimates for a given set of observables %
which are not necessarily informationally complete~\cite{R68,R69,R70,R71,R72}. In this setting, our estimate of the unknown quantum state is the one that maximizes the entropy subject to the constraints corresponding to our expectation-value estimates.
Recent years have seen a resurgence in the idea of learning using incomplete
knowledge via the topic of \emph{shadow tomography}, %
i.e., learning a
state in terms of its expectation values on a given set of observables, often
provided randomly from a known
ensemble~\cite{R57,R88}. This concept has been
combined with the maximum-entropy principle to obtain quantum state learning algorithms~\cite{R73,R74,R59,R60}, including for Choi states of
  processes~\cite{R89}.

Here, we extend this idea to quantum channels, capturing the full channel nature of
the learning task
by using the quantum channel relative entropy.  %
We consider an online learning setting in which we are tasked with learning a quantum channel in a sequential manner. Starting with the completely-depolarizing channel $\mathcal{D}$ as our initial guess, at each iteration, our
learning algorithm (see \cref{z:igCtH9MB3u3E}) estimates the expectation value of a given channel observable by making use the unknown channel a
fixed number of times. The estimate incurs a loss, depending how close it is to the true expectation value, and this loss is used to compute an
updated estimate of the unknown channel. This algorithm is a direct generalization of the
learning algorithm considered in refs.~\cite{R90,R73} in the context of quantum state learning. 

\begin{algorithm}[H]
  \begin{algorithmic}[1]
    \Require {$\eta \in (0, 1)$; $\mathcal{M}^{(0)} = \mathcal{D}$.}
    \For{\texttt{$t = 1, 2, \ldots, T$}}
    \State Receive the observable $E^{(t)}$.
    \State Obtain an estimate $s^{(t)}$ of the true expectation value.
    \State Update: $\mathcal{M}^{(t)}=\text{argmin}_{\mathcal{N}\text{ cp. tp.}} {D}(\mathcal{N}\mathclose{}\,\Vert\,\mathopen{}\mathcal{M}^{(t-1)}) + \eta L_t(\mathcal{N})$.
    \EndFor
    \Ensure { $\mathcal{M}^{(T)}$} 
  \end{algorithmic}
  \caption{Minimum relative entropy channel learning}
  \label{z:igCtH9MB3u3E}
\end{algorithm}

\begin{figure}
    \centering
    \includegraphics{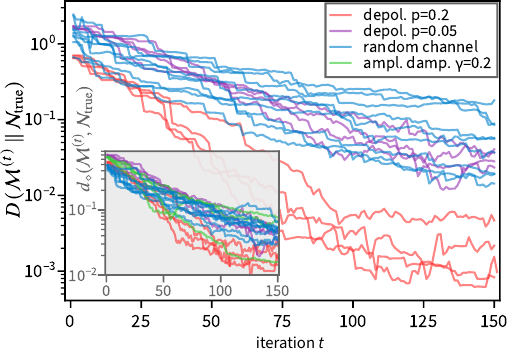}
    \caption{Simulated runs of \Cref{z:igCtH9MB3u3E} to learn a depolarizing channel
      (with $p=0.2$ and $p=0.05$; $4$ runs each), an amplitude-damping channel
      ($4$ runs), and $9$ random channels. At each iteration $t$, the algorithm
      updates an estimate $\mathcal{M}^{(t)}$ of the unknown channel
      $\mathcal{N}_{\mathrm{true}}$ by solving a minimum channel relative
      entropy problem that involves a new estimated expectation value $s^{(t)}$
      of a channel observable $E_{BR}^{(t)}$.  
      The channel relative
      entropy ${D}(\mathcal{M}^{(t)}\mathclose{}\,\Vert\,\mathopen{}\mathcal{N}_{\mathrm{true}})$ appears to decay
      towards zero, indicating the estimate approaches the true channel.
      \emph{Inset:} $\mathcal{M}^{(t)}$ also appears to approach
      $\mathcal{N}_{\mathrm{true}}$ in the diamond distance
      $d_\diamond(\mathcal{M}^{(t)}, \mathcal{N}_{\mathrm{true}}) = (1/2)\lVert {
        \mathcal{M}^{(t)} - \mathcal{N}_{\mathrm{true}} }\rVert _\diamond$.  The runs with the
      amplitude damping channel appear only in the diamond distance plot since
      the relative channel entropy is not well defined in this case.  Our
      numerics employ the techniques of
      refs.~\protect\cite{R91,R83}.}
    \label{z:6osEYsZSns92}
\end{figure}

The loss function is $L_t(\mathcal{N})\coloneqq\bigl({s^{(t)}-\operatorname{tr}[E^{(t)}N]}\bigr)^2$, where $N$ is the Choi representation of $\mathcal{N}$. In our numerical implementation of \Cref{z:igCtH9MB3u3E}, the observables are of the form $E=P\otimes\rho$, where $P\in\{X,Y,Z\}$ is a non-identity Pauli operator and
$\rho\in\{\lvert {0}\rangle \mkern -1.8mu\relax \langle{0}\rvert ,\lvert {1}\rangle \mkern -1.8mu\relax \langle{1}\rvert ,\lvert {\pm}\rangle \mkern -1.8mu\relax \langle{\pm}\rvert ,\lvert {\pm i}\rangle \mkern -1.8mu\relax \langle{\pm i}\rvert \}$ is a single-qubit stabilizer state. In every iteration of \Cref{z:igCtH9MB3u3E}, we make a uniformly random choice of
$P\in\mathcal{P}$ and $\rho\in\mathcal{S}$, and set %
$\eta=0.15$. The quantity $\eta$ is a \emph{learning rate}, which models the tradeoff between keeping the new channel estimate close to the old one, represented by the first term in the objective function of the update step, and minimizing the loss in the second term. The estimate $s^{(t)}$ is obtained via measurement of the given observable and aggregating the results with previous measurement outcomes of the same observables~\cite{R73}. Our results are presented in Fig.~\ref{z:6osEYsZSns92}.

Our study represents a simple proof of concept of the performance of
\cref{z:igCtH9MB3u3E} for a selection
of single-qubit channels, leaving rigorous convergence guarantees and
demonstration of learning of larger-scale channels
beyond the scope of this work.

\paragraph{Methods and main technical advances.}
Our methods involve mathematically rigorous proofs based on modern convex
optimization tools, quantum typicality techniques for quantum states and
channels~\cite{R92,R93,R86,R94,R77}, as well as Schur-Weyl duality~\cite{R95,R96}.
We single out two technical advances of our work here (cf.\@ details in our
companion paper~\cite{R50}).  First, we prove a new
postselection theorem, extending those of
refs.~\cite{R97,R98,R99,R100,R101}, which operator-upper-bounds any
permutation-invariant quantum channel $\mathcal{E}$ by a mixture of i.i.d.\@
channels $\mathcal{M}^{\otimes n}$ weighted by the proximity of
$\mathcal{M}^{\otimes n}$ to $\mathcal{E}$.  This technical tool is useful to
prove properties for an arbitrary $\mathcal{E}$ that are more easily proven for
an i.i.d.\@ channel $\mathcal{M}^{\otimes n}$ (such as the concentration of
outcomes of $\overline{H^{j,\sigma}}_{B^nR^n}$).
Second, our approach formalizes a new concept of typicality for quantum
channels.
Typicality is key technical tool for proving coding theorems in classical and
quantum information theory (among others)~\cite{R102,R103,R92,R104};
this concept has seen various extensions for
channels~\cite{R105,R106,R107,R108,R109,R77}.  Here, our microcanonical operator $P_{B^nR^n}$ defines a
mathematical object analogous to a quantum state's \emph{typical projector}, in
that it selects $n$-copy channels with certain statistical concentration
properties.
(Simpler attempts to define such operators tend to fail; for instance, a typical
or microcanonical projector for a channel's Choi state
$[\mathcal{M}(d_A^{-1}\Phi_{A:R})]^{\otimes n}$ would fail to attribute high
weight to operators of the type $[\mathcal{M}(\sigma_{AR})]^{\otimes n}$ for
non-maximally-mixed input states $\sigma_R$.)

\paragraph{Discussion.}
We extend two fundamental principles that define the thermal state to quantum
channels, Jaynes' maximum entropy principle and the microcanonical approach, and
prove that they lead to the same thermal quantum channel.  The thermal quantum
channel's further application in the task of learning an unknown processes
suggests that the thermal quantum channel's uses extend significantly beyond the
setting of thermalizing systems, much like the thermal quantum state appears in
quantum state inference algorithms.

By modeling the dynamics rather than the final state of a system, the thermal
quantum channel can model systems %
that are only partially thermalizing or which
keep some memory of the input state.
For example, the average energy conservation constraint example discussed above
implements a form of ``local relaxation'' whereby the average energy is
conserved, but the output state is consistently a canonical Gibbs state for any
input energy eigenstate.  Local relaxations after quenches have been studied in
the contexts of Gaussian systems with clustering correlations and central limit
theorems~\cite{R110,R111,R112}.  The thermal quantum channel might provide a
consistent, general-purpose approach to describe the local thermalizing
projection dynamics of such systems.
We also anticipate that the thermal quantum channel
can model settings with multiple thermalization mechanisms or
with a separation of different relaxation time scales, including regimes of
hydrodynamic behavior~\cite{R19,R113,R24}.

The maximum channel entropy principle interprets thermalizing dynamics
as a channel that attempts to increase the systems' entropy as
much as possible, while remaining compatible with the constraints.
The maximum channel entropy principle assumes that there are no further
obstacles to thermalization other than any explicitly stated linear constraints.
Importantly, the principle cannot be invoked to make any statements about
whether $\mathcal{U}$ thermalizes in the first place. %

Our $\mathcal{T}$ does not generally have a Choi state of thermal form.
Rather, a channel with a thermal Choi state would
maximize the channel's output entropy (relative to $R$) only for the maximally
entangled input state between $A$ and $R$; such a channel might still produce
low-entropy outputs for other input states.

Our generalized minimum channel relative entropy problem (cf.~\cite{R50}) includes inequality constraints, a
quadratic loss function, and optimization relative to a reference channel.
Furthermore, our setting can express constraints beyond the examples considered above.
For instance, $\mathcal{T}$ can be constrained within a fixed-sized light cone by
demanding that correlation functions for faraway sites vanish; or $\mathcal{T}$
might model a short-time evolution by demanding that
$\operatorname{tr}[{\Phi_{B:R} \mathcal{T}(\Phi_{A:R})}]/d_R \geq 1-\epsilon$.

The thermal quantum channel can be numerically approximated with semidefinite programming
methods~\cite{R91,R83}.  Yet known
difficulties for computing properties for the thermal state are inherited in our case;
indeed, the thermal quantum state is a special case of the thermal quantum
channel with a trivial input system.  Such difficulties include determining
the $\mu_j$'s or computing the thermal state's
partition function~\cite{R114}.
Our work supports %
 exciting prospects for extending to channels a broad literature of
modern techniques for studying thermal states,
including tensor networks (e.g.,~\cite{R115,R116}
and references therein),
information-theoretic bounds and decay of correlations (e.g.,~\cite{R117,R118,R119,R120}), as well as
quantum algorithms (e.g.,~\cite{R121,R122,R123}).

We expect that further approaches to characterize the thermal state could
be extended to quantum channels, such as complete passivity~\cite{R124,R125,R86},
via its role in the resource theory of
thermodynamics~\cite{R126,R127,R128,R129,R130,R131,R132}, and canonical typicality~\cite{R7}.
Furthermore, we expect %
 certain systems could be proven to
equilibrate dynamically over time to the thermal channel,
extending thermalization results for
states~\cite{R133,R134,R135,R9,R10,R11,R136}.

Our results enrich the picture of thermalization in physics by viewing it as
a full quantum process, rather than the equilibration to a fixed, final state,
and we provide a robust theoretical foundation to model thermalizing processes
that conserve memory of the initial state.
Furthermore, the widespread relevance of the canonical Gibbs state throughout
information theory, quantum thermodynamics, machine learning, and quantum algorithms
provides a promising outlook for similar applications of the quantum thermal channel.

\paragraph{Acknowledgments}
The authors thank
David Jennings,
Jens Eisert,
Michael Walter,
Hamza Fawzi, Omar Fawzi,
Gereon Ko\ss{}mann,
Mark Mitchison,
John Preskill,
and Andreas Winter
for insightful discussions.
PhF is supported by the project FOR 2724 of the Deutsche Forschungsgemeinschaft
(DFG).

\paragraph{Note added}
Our results were submitted to
\href{https://sites.google.com/view/beyondiid13}{%
  \emph{Beyond i.i.d.\@ in information theory} 2025} in April 2025
and accepted as a talk in early June 2025 (cf.\@ 
\url{https://sites.google.com/view/beyondiid13/program}).
During the final stages of completion of our manuscript,
a paper with independent related work by Siddhartha Das and Ujjwal Sen
appeared on the arXiv on July 1, 2025 [Das and Sen, arXiv:2506.24079].

\clearpage
\newgeometry{hmargin=1.5in,vmargin=1.25in}
\onecolumngrid
\setstretch{1.1}
\appendix

\section{Appendix: Technical theorem statements}

We present technical statements of our main results for completeness and to
ensure a self-contained presentation of the technical results underlying this
paper.
Our technical proofs are detailed and discussed in the dedicated companion
paper~\cite{R50}.  

The maximum-channel entropy problem is stated as follows:
\begin{align}
  \label{z:LSu6pAIaay5c}
  \begin{aligned}[t]
    \textup{maximize:} \quad
    & {S}(\mathcal{N}_{A\to B})
    \\
    \textup{over:}\quad
    & \mathcal{N}_{A\to B}\ \textup{completely positive, trace-preserving map}
    \\
    \textup{such that:}\quad
    & \operatorname{tr}\bigl[{C^j_{BR}\,\mathcal{N}_{A\to B}(\Phi_{A:R})}\bigr] = q_j\quad\text{for \(j=1, \ldots, J\)}\ .
  \end{aligned}
\end{align}

\begin{theorem*}[Structure of the thermal channel]
  A quantum channel $\mathcal{T}_{A\to B}$ is optimal
  in~\eqref{z:LSu6pAIaay5c} if and only if it satisfies all the
  problem constraints and it has a Choi matrix of the form
  \begin{align}
    \mathcal{T}_{A\to B}(\Phi_{A:R})
    = \phi_R^{-1/2}
    {e}^{
    - \phi_R^{-1/2}\Bigl[{
    \sum \mu_j C^j_{BR}
    -\mathds{1}_B\otimes ({F_R + \phi_R\log\phi_R}) 
    - S_{BR} }\Bigr] \phi_R^{-1/2}
    }
    \phi_R^{-1/2}
    + Y_{BR}\ ,
  \end{align}
  where:
  \begin{itemize}[parsep=0pt]
    \item $\mu_j\in\mathbb{R}$, $j=1,\ldots, J$;
    \item $F_R$ is a Hermitian operator; 
    \item $S_{BR}$ is a positive semidefinite operator satisfying
      $S_{BR}\,\mathcal{T}_{A\to B}(\Phi_{A:R}) = 0$;
    \item it holds that
      $\Pi_R^{\phi_R\perp}\bigl({\sum \mu_j C^j_{BR}
      -\mathds{1}_B\otimes F_B  - S_{BR} }\bigr) = 0$;
    \item $Y_{BR}$ is a
      Hermitian operator satisfying $\Pi^{\phi_R}_R Y_{BR} \Pi^{\phi_R}_R = 0$; and
    \item $\phi_R$ is the local reduced state on $R$ of an optimal state
      $\lvert {\phi}\rangle _{AR} = \phi_R^{1/2}\lvert {\Phi_{A:R}}\rangle  =
      \phi_A^{1/2}\lvert {\Phi_{A:R}}\rangle $ in the definition of the channel entropy
      ${S}(\mathcal{T}_{A\to B}) = \min_{\lvert {\phi}\rangle _{AR}} {S}(B\mathclose{}\,|\,\mathopen{}R)_{{\mathcal{T}_{A\to
          B}({\phi_{AR}})}}$.
  \end{itemize}
  The channel entropy attained by $\mathcal{T}_{A\to B}$ is
  \begin{align}
    {S}(\mathcal{T}_{A\to B}) = -\operatorname{tr}({F_R}) + \sum_{j=1}^J \mu_j q_j\ .
  \end{align}
  Furthermore, any optimal state $\phi_A$ (with
  $\phi_A = \operatorname{tr}_R({\phi_{AR}}) = \phi_R^{t_{R\to A}}$) must satisfy
  \begin{align}
    \log(\phi_A) - \widehat{\mathcal{T}}^\dagger\Bigl({
    \log\bigl[{ \widehat{\mathcal{T}}_{A\to E}({\phi_A}) }\bigr] }\Bigr) \ \propto\  \Pi_A\ ,
    \label{z:lWRPNfFIPCpB}
  \end{align}
  where $\widehat{\mathcal{T}}_{A\to E}$ is a complementary channel to
  $\mathcal{T}_{A\to B}$.  If $\phi_A$ has full rank, then
  $S_{BR} = 0 = Y_{BR}$, and~\eqref{z:lWRPNfFIPCpB}
  is sufficient for optimality of $\phi_A$.  
\end{theorem*}

If $\phi_R$ has full rank, then $Y_{BR} = 0$, $S_{BR} = 0$, and $\mathcal{T}_{A\to B}$
is unique.  
Further theorem statements also require the notion of a \emph{thermal quantum
channel with respect to a fixed state $\phi_R$}.  Let $\phi_R$ be any quantum state.
A \emph{thermal quantum channel with respect to $\phi_R$}, denoted by
$\mathcal{T}^{(\phi_R)}$, is an optimal solution to the following problem:
\begin{align}
  \label{z:S65dJbbWg2cN}
  \begin{aligned}[t]
    \textup{maximize:} \quad
    & {S}(B\mathclose{}\,|\,\mathopen{}R)_{{\mathcal{N}_{A\to B}\bigl({\phi_R^{1/2}\Phi_{A:R}\phi_R^{1/2}}\bigr)}}
    \\
    \textup{over:}\quad
    & \mathcal{N}_{A\to B}\ \textup{completely positive, trace-preserving map}
    \\
    \textup{such that:}\quad
    & \operatorname{tr}\bigl[{C^j_{BR}\,\mathcal{N}_{A\to B}(\Phi_{A:R})}\bigr] = q_j
      \quad\text{for \(j=1, \ldots, J\)}\ .
  \end{aligned}
\end{align}
If we minimize the resulting objective over $\phi_R$, we obtain the
problem~\eqref{z:LSu6pAIaay5c}.  (See
ref.~\cite{R50} for details.)
If $\phi_R$ has full rank, the optimizer $\mathcal{T}^{(\phi_R)}$ is unique and
is a continuous function on the set of full-rank states $\phi_R$.  
Furthermore, if
$(\phi_R^z)_{z>0}$ is a family of states with
$\phi_R \equiv \lim_{z\to0}\phi_R^z$ and
$\mathcal{T} \equiv \lim_{z\to0} \mathcal{T}^{(\phi_R^z)}$, then $\mathcal{T}$
is a thermal quantum channel with respect to $\phi_R$.

\begin{definition*}[Approximate microcanonical channel operator]
  An operator $P_{B^nR^n}$ satisfying $0\leq P \leq \mathds{1}$ is called an
  \emph{$(\eta,\epsilon,\delta, y, \nu, \eta',\epsilon',\delta', y',
    \nu')$-approximate microcanonical channel operator} with respect to
  $\{{ (C^j_{BR}, q_j) }\}$ if the following two conditions hold.  The conditions
  are formulated in terms of
  $P_{B^nR^n}^\perp \equiv \mathds{1}_{B^nR^n} - P_{B^nR^n}$ and use the shorthand
  $\lvert {\sigma_{AR}}\rangle  \equiv \sigma_R^{1/2}\lvert {\Phi_{A:R}}\rangle $ for any $\sigma_R$:
  \begin{enumerate}[label=(\alph*)]
  \item For any channel $\mathcal{E}_{A^n\to B^n}$ such that
    \begin{align}
      \max_{\sigma_{R}\geq y\mathds{1}} \operatorname{tr}\bigl[{
      P_{B^n R^n}^\perp \, \mathcal{E}_{A^n\to B^n}\bigl({\sigma_{AR}^{\otimes n}}\bigr)
      }\bigr] &\leq \epsilon\ ,
    \end{align}
    then for all $j=1, \ldots, J$,
    \begin{align}
      \max_{\sigma_{R}\geq \nu y\mathds{1}} \operatorname{tr}\Bigl[{
        \Bigl\{{ \overline{H^{j,\sigma}}_{B^nR^n} \notin [q_j\pm \eta ] }\Bigr\} \,
        \mathcal{E}_{A^n\to B^n}(\sigma_{AR}^{\otimes n})
        }\Bigr] &\leq \delta\ ,
        \label{z:APDcifUuNl36}
    \end{align}
    where $\bigl\{{ X \notin I }\bigr\}$ denotes the projector onto the eigenspaces of a
    Hermitian operator $X$ associated with eigenvalues not in a set
    $I\subset\mathbb{R}$.

  \item %
    For any channel $\mathcal{E}_{A^n\to B^n}$ such that
    \begin{align}
      \max_{\sigma_{R}\geq y'\mathds{1}} \operatorname{tr}\Bigl[{
        \Bigl\{{ \overline{H^{j,\sigma}}_{B^nR^n} \notin [q_j\pm \eta' ] }\Bigr\} \,
        \mathcal{E}_{A^n\to B^n}(\sigma_{AR}^{\otimes n})
        }\Bigr] &\leq \delta' \qquad\text{for all}\ j=1,\ldots, J\ ,
    \end{align}
    then
    \begin{align}
      \max_{\sigma_{R}\geq \nu' y'\mathds{1}} \operatorname{tr}\bigl[{
      P_{B^nR^n}^\perp \mathcal{E}_{A^n\to B^n}(\sigma_{AR}^{\otimes n})
      }\bigr] &\leq \epsilon'\ .
          \label{z:msGKg5m1gifg}
    \end{align}
  \end{enumerate}
\end{definition*}

\begin{definition*}
  Let $P_{B^nR^n}$ be a
  $(\eta, \epsilon, \delta, y, \nu, \eta', \epsilon', \delta', y',
  \nu')$-approximate microcanonical channel operator with respect to
  $\{{ (C^j_{BR}, q_j) }\}$.  Then the associated \emph{approximate microcanonical
    channel} is defined as the channel $\Omega_{A^n\to B^n}$ that maximizes the
  channel entropy ${S}(\Omega_{A^n\to B^n})$ subject to the constraint
  \begin{align}
    \max_{\sigma_{R}\geq y\mathds{1}} \operatorname{tr}\bigl[{
        P^\perp_{B^nR^n} \, \Omega_n\bigl({
            \sigma_{AR}^{\otimes n}
        }\bigr)
    }\bigr]
    \leq \epsilon\ .
    \label{z:Y7YnuhFNVrMt}
  \end{align}
\end{definition*}

\begin{theorem*}[The microcanonical channel resembles the thermal channel on a single copy]
  Let $\Omega_n$ be a approximate microcanonical channel associated with a
  $(\eta,\epsilon,\delta,y,\nu, \eta',\epsilon',\delta',y',\nu')$-approximate
  microcanonical channel operator $P_{B^nR^n}$, and let
  \begin{align}
    \omega_{BR}
    = \frac1n \sum_{i=1}^n\operatorname{tr}_{n\setminus i}\bigl[{\Omega_n\bigl({\phi_{AR}^{\otimes n}}\bigr)}\bigr]\ ,
  \end{align}
  where $\operatorname{tr}_{n\setminus i}$ denotes the partial trace over all copies of $(BR)$
  except $(BR)_i$, and where $\phi_R$ is any full-rank state with
  $\lambda_{\mathrm{min}}(\phi_R) \geq \nu y$ and
  $\lambda_{\mathrm{min}}(\phi_R) \geq y'$.  Let $\mathcal{T}_{A\to B}^{(\phi)}$
  be the thermal channel with respect to $\phi$.
  Assume that $2\lVert {C^j_{BR}}\rVert ^2\,\log({2/\delta'}) \leq n \eta'^2 y'^2$ for
  all $j=1,\ldots,J$.  Additionally, we assume that $\epsilon' \leq \epsilon$.
  Then
  \begin{align}
    {D}\bigl ( \omega_{BR} \mathclose{}\,\big \Vert\,\mathopen{}
    \mathcal{N}_{\mathrm{th}}(\phi_{AR}) \bigr )
    \leq
    \sum \mu_j \bigl({\eta + 2y^{-1}\bigl \lVert {C^j_{BR}}\bigr \rVert \,\epsilon}\bigr)\ ,
  \end{align}
  where ${D}(\rho\mathclose{}\,\Vert\,\mathopen{}\sigma) = \operatorname{tr}\bigl({\rho\bigl[{\log({\rho})-\log({\sigma})}\bigr]}\bigr)$ is
  the Umegaki quantum relative entropy.
\end{theorem*}

Let $\mathcal{T}$ and $\phi_R$ be optimal
in~\eqref{z:LSu6pAIaay5c} for the same constraints as in the
construction of $P_{B^nR^n}$, such that $\mathcal{T}$ is a limit of thermal
quantum channels with respect to $\phi_R^{(n)}$ with
$\phi_R^{(n)}\geq \max\{{\nu y(n), y'(n)}\}\,\mathds{1}$ and where (say)
$y(n) = y'(n) = 1/n^{0.01}$ and $\nu=\nu'=3/2$ (see parameter regimes below).
The above theorem then implies that the single-copy effective process of the
microcanonical channel on $n$ copies resembles $\mathcal{T}$ in the limit of
large $n$, by ensuring that $\omega_{BR}^{(n)}$ resembles
$\mathcal{T}^{(\phi_R^{(n)})}$, which itself converges towards $\mathcal{T}$.

\begin{theorem*}[Existence of an approximate microcanonical channel operator]
  Let $\boldsymbol q = \{{ q_j }\}_{j=1}^J$, let
  $0<\eta'<\eta < \min_j\lVert {C^j_{BR}}\rVert $, and write $\bar\eta=(\eta'+\eta)/2$.
  There exists a two-outcome POVM $\{{ P_{B^nR^n}, P^\perp_{B^nR^n} }\}$ such that:
  \begin{enumerate}[(\roman*)]
  \item \label{z:W3JG7egy3IKB}%
    For any $\epsilon>0$, $\nu>1$, and for any $0<y<1/(\nu d_R)$, let
    $\mathcal{E}_{A^n\to B^n}$ be any quantum channel such that
    \begin{align}
      \max_{\sigma_R\geq y\mathds{1}}
      \operatorname{tr}\bigl[{ P^\perp_{B^nR^n} \mathcal{E}\bigl({\sigma_{AR}^{\otimes n}}\bigr) }\bigr]
      \leq \epsilon\ ,
    \end{align}
    using the shorthand $\lvert {\sigma}\rangle _{AR} \equiv \sigma_R^{1/2}\lvert {\Phi_{A:R}}\rangle $.
    Assume furthermore that
    $\nu \geq 1+ ({\eta-\eta'})/(4\max_j \lVert {C^j_{BR}}\rVert )$.  Then, for any
    $j=1, \ldots, J$,
    \begin{multline}
      \max_{\sigma_R\geq \nu y\mathds{1}} \operatorname{tr}\Bigl[{
      \bigl\{{ \overline{H^{j,\sigma}}_{B^nR^n} \notin [{q_j \pm \eta}] }\bigr\}
      \,
      \mathcal{E}_{A^n\to B^n}\bigl({\sigma_{AR}^{\otimes n}}\bigr)
      }\Bigr]
      \\
      \leq \operatorname{poly}({n}) \exp\Biggl\{{
        -n y^8 \min\biggl({
            -\frac{\log(\epsilon)}{n y^8}
            \,,\;
            \frac{c'({\eta-\eta'})^8}{\max_j \lVert {C^j_{BR}}\rVert ^8}
        }\biggr)
      }\Biggr\}\ ,
      \label{z:jFwlnxgTFNKU}
    \end{multline}
    with $c'=1/(2\times 5^8)$.

  \item \label{z:wEDQuyLiMLq1}%
    For any $\delta'>0$, $\nu' > 1$, and for any
    $0<y'<1/(\nu' d_R)$, let $\mathcal{E}_{A^n\to B^n}$ be any quantum channel
    such that for all $j=1, \ldots, J$,
    \begin{align}
      \max_{\sigma_R\geq  y'\mathds{1}} \operatorname{tr}\Bigl[{
      \bigl\{{ \overline{H^{j,\sigma}}_{B^nR^n} \notin [{q_j \pm \eta'}] }\bigr\}
      \,
      \mathcal{E}_{A^n\to B^n}\bigl({\sigma_{AR}^{\otimes n}}\bigr)
      }\Bigr]
      \leq \delta'\ ,
    \end{align}
    using the shorthand $\lvert {\sigma}\rangle _{AR} \equiv \sigma_R^{1/2}\lvert {\Phi_{A:R}}\rangle $.
    Assume furthermore that
    $\nu' \geq 1+ ({\eta - \eta'})/(4 \max_j\lVert {C^j_{BR}}\rVert )$.  Then
    \begin{align}
      \max_{\sigma_R\geq \nu' y'\mathds{1}}
      \operatorname{tr}\bigl[{ P^\perp_{B^nR^n} \mathcal{E}\bigl({\sigma_{AR}^{\otimes n}}\bigr) }\bigr]
      &
      \leq
      \operatorname{poly}({n}) \exp\Biggl\{{
      -n y'^8 \min\Biggl({
      -\frac{\log\bigl({\delta'}\bigr)}{ n y'^8 } ,
      \frac{c' ({\eta-\eta'})^8}{\,\max_j \lVert {C^j_{BR}}\rVert ^8 }
      }\Biggr)
      }\Biggr\}
      \ ,
      \label{z:aM9IqJnTH1Z8}
    \end{align}
    with $c'=1/(2\times 5^8)$.
  \end{enumerate}
\end{theorem*}

Parameter regimes in which the above theorems are successful include, for large
enough $n$:
\begin{align}
    y &= n^{-\beta_1}\ ;
  &
    y' &= n^{-\beta_2}\ ;
  &
    \eta &= c_{\mathrm{min}} n^{-\gamma}\ ;
  &
    \eta'&= \eta/2\ ;
  &
    \nu &= \nu' = 3/2\ ,
\end{align}
with $c_{\mathrm{min}} \equiv \min_j \lVert {C^j_{BR}}\rVert $,
$0< \gamma=\beta_1=\beta_2 < 1/16$.  These parameters lead to
\begin{align}
    \begin{aligned}
  \epsilon &= \exp\bigl({-n^{1-17\gamma}}\bigr)\ ;
  &\qquad\qquad
  \delta &= \operatorname{poly}({n}) \exp\bigl({ - n^{1-17\gamma} }\bigr)\ ;
  \\[1ex]
    \delta' &= \exp\bigl({-n^{1-5\gamma}}\bigr)\ ;
  &
    \epsilon' &= \operatorname{poly}({n}) \exp\bigl({ - n^{1-17\gamma} }\bigr) \ .
  \end{aligned}
\end{align}

\end{document}